\documentclass[amssymb,11pt]{article}
\usepackage[pdftex,colorlinks=true,urlcolor=black,filecolor=black,linkcolor=black,
            pdftitle={The clebsch momenta approach to fluid quantization.},
            pdfauthor={Mark D. Roberts},
            pdfsubject={Fluid Quantization},
            pdfkeywords={Fluid Hamiltonian,  quantization},
            pagebackref,pdfpagemode=None,bookmarksopen=true]{hyperref}
\usepackage{amssymb}
\usepackage{graphicx}
\newcommand{\bc}{\begin{center}}
\newcommand{\ec}{\end{center}}
\newcommand{\be}{\begin{equation}}
\newcommand{\ee}{\end{equation}}
\newcommand{\ber}{\begin{eqnarray}}
\newcommand{\ear}{\end{eqnarray}}

\newcommand{\bx}{\Box}

\newcommand{\Hm}{{\cal H}}

\newcommand{\Lg}{{\cal L}}
\newcommand{\n}{\nonumber\\}

\newcommand{\pr}{{\cal P}}

\newcommand{\st}{\stackrel}

\newcommand{\p}{\partial}
\begin{document}
\title{The clebsch momenta approach to fluid lagrangians.}
\author{
\href{http://www.violinist.com/directory/bio.cfm?member=robemark}
{Mark D. Roberts},\\
54 Grantley Avenue,  Wonersh Park,  GU5 0QN,  UK\\
mdr@ihes.fr
}
\date{$18^{th}$ of October 2009}
\maketitle
\begin{abstract}
The clebsch potential approach to fluid lagrangians is developed in order to establish
contact with other approaches to fluids.
Three variants of the perfect fluid approach are looked at.
The first is an explicit linear lagrangian constructed directly from the clebsch potentials,
this has fixed equation of state and explicit expression for the pressure but is less general than a perfect fluid.
The second is lagrangians more general than that of a perfect fluid which are constructed from
higher powers of the comoving vector.
The third is lagrangians depending on two vector fields which can represent both density flow
and entropy flow.
\end{abstract}
{\scriptsize \tableofcontents}
\section{Introduction}\label{intro}
\subsection{Motivation}\label{motivation}
The motivation of this paper is to provided a unified approach to the various ways that fluids are described in physics.
In particular the methods used by relativists,  fluid mechanists,  and nuclear physicists have grown distinct.
In many areas of physics a unified approach is provided by the lagrange method,
which for fluids is developed here.
\subsection{Methodology}\label{methodology}
The methodolgy used is first to simplify a perfect fluid in order to investigate if methods
of field theory can be applied to it;
and then to generalize a perfect fluid to try and establish contact with more physical fluids.
\subsection{Other approaches to fluids}\label{contact}
A perfect fluid has a variational formulation \cite{hargreaves,RWH,SW,taub}
which uses the first law of thermodynamics.
In such a formulation clebsch potentials \cite{clebsch,lamb,baldwin,balkovsky,GH,rund79,eckart}
for the comoving fluid vector field are used.
Here this approach is both applied to less general fluids and to more general fluids.
Other approaches to fluids include the following {\bf twelve}.
The {\it first} uses lagragians dependent on combinations of clebsh potentials
which do not necessarily form a vector \cite{mdr41}.
The {\it second} is that the comoving vector can be thought of as $U^a=\dot{x}^a$,
so that a perfect fluid is a type of generalization of a point particle,
then there turns out to be a fluid generalization of a membrane \cite{mdrfm}.
The {\it third} is that the charge subsitution $\p_a\rightarrow\p_a+\imath eA_a$
can be applied to fluids as well as fields
and this leads to a model of symmetry breaking \cite{mdr25}.
The {\it fourth} is that the navier-stokes equation has a lagrangian formulation
\cite{CBB,constantin,DMB,finlayson,UC},
but the lagrangian has different measure and also image fields.
The {\it fifth} is that hydrodynamics can be expressed using a
grad expansion \cite{grad,IS,HL,EXG} which needs an entropy vector.
The {\it sixth} is that contemporary bjorken models
use the grad expansion \cite{bjorken,moss,muronga,NS,HJP}.
The {\it seventh} is fluid plasmas \cite{anile,achterberg}.
The {\it eighth} is elastic models \cite{ABS}\S3,
where the density rather than the pressure is used as the lagrangian.
The {\it ninth} is other quantization methods such as brst and path integral
applied to fluids \cite{BH}.
The {\it tenth} is superfluids \cite{CK}.
The {\it eleventh} is spinning fluids \cite{HHKN,jackiw,NHN}.
The {\it twelfth} is cosmology \cite{LR},  where clebsch potentials have been used.
\subsection{Conventions}\label{conventions}
The word potential is disambiguated by refereing to potentials for a vector field as clebsch
potentials and potentials that occur in lagrange theory as coefficient functions.
When a measure is suppressed it is $\int\sqrt{-g}dx^4$ not $\int d\tau$ unless otherwise stated.
$\mu$ is density and $\pr$ is the pressure.  $q$ is a clebsch potential.
$\sigma$ is used for a clebsch potential,  a pauli matrix and the shear of a vector,
to disambiguate the pauli matrix is always $\sigma_p$ and the shear labels with which vector
it is with respect to $\st{U}{\sigma}$.
Capital $\Pi$ indicates a momentum with respect to the proper time $\tau$
not the coordinate time $t$.
$a,b,c,\ldots$ are spacetime indices,
$i,j,k,\ldots$ label sets of fields and momenta,  and
$\iota,\kappa,\ldots$ label constraints.
The signature is $-+++$.
\section{The perfect fluid}\label{pf}
For a perfect fluid the lagrangian is taken to be the pressure $\Lg=\pr$,  and the action is
\be
I=\int dx^4\pr.
\label{pfaction}
\ee
The clebsch potentials are given by
\be
hV_a=W_a=\sigma_a+\theta s_a,~~~~~~~
V_aV^a=-1,
\label{simplecleb}
\ee
where if more potentials are needed it is straightforward to instate them;
there are several sign conventions for \ref{simplecleb}.
The clebsch potentials are sometimes given names:
$\sigma$ is called the higgs because it has a similar role to the higgs field in symmetry
breaking using fluids \cite{mdr25,mdr41},
$\theta$ is called the thermasy and $s$ the entropy \cite{SW}.
Variation is achieved via the first law of thermodynamics
\be
\delta\pr=n\delta h-nT\delta s=-nV_a\delta W^a-nT\delta s,~~~~~~~
nh=\mu+\pr,
\label{1law}
\ee
where $n$ is the particle number and $h$ is the enthalpy.
Metrical variation yields the stress
\be
T_{ab}=(\mu+\pr)V_aV_b+\pr g_{ab},
\label{pfstress}
\ee
the n\"other currents $j^a=\delta I/\delta q_a$ are
\be
j_\sigma^a=-nV^a,~~~
j_\theta^a=0,~~~
j_s^a=-n\theta V^a,
\label{noether}
\ee
variation with respect to the clebsch potentials gives
\be
(nV^a)_a=\dot{n}+n\Theta=0,~~~
\dot{s}=0,~~~
\dot{\theta}=T,
\label{cbemo}
\ee
where $\Theta\equiv V^a_{.;a}$ is the vectors expansion:
thus the consevation of the n\"other currents \ref{noether}
gives the same equations \ref{cbemo} as varying the clebsch potentials;
the normalization condition $V_aV^a=-1$ and \ref{cbemo} give
\be
\dot{\sigma}=-h.
\label{shpf}
\ee
The bianchi identity is
\be
T^{ab}_{..;b}=n\dot{W}^a+\pr^{,a},
\label{bid}
\ee
subsituting for $W$ using \ref{simplecleb} and for $\pr$ using \ref{1law}
this vanishes identically.
If one attempts to apply existing scalar field fourier oscillator quantization procedures
to the above there is the equation
\be
W^a_{.;a}
=\bx\sigma+\theta_as^a+\theta\bx s
=\left(hV^a\right)_a
=\dot{h}+h\Theta
=\dot{h}-\frac{\dot{n}}{n}h
=h\left(\ln\left(\frac{h}{n}\right)\right)^\circ,
\label{pfbxsig}
\ee
and if this vanishes the enthalpy $n$ is proportional to the particle number $n$,
for an example of the see the next section \S\ref{eospf}.
The pressure $\pr$ and density $\mu$ are only implicity
defined in terms of the clebsch potentials so it is not clear
what operators should correspond to them.
Another possibility is to note that \ref{cbemo} are first order differential equations
and to try and replace them with spinorial equations;
however this would require a spinorial absolute derivative
in place of the vectorial absolute derivative,  see \cite{PR}\S4.4.

The canonical clebsch momenta are given by $\Pi^i=\delta{I}/\delta\dot{q^i}$
\be
\Pi^\sigma=-n,~~~
\Pi^\theta=0,~~~
\Pi^s=-n\theta,
\label{pfmom}
\ee
and these allow the n\"other currents \ref{noether} to be expressed as
\be
j^a_q=\Pi_qV^a.
\label{jpi}
\ee
The standard poisson bracket is defined by
\be
\left\{A,B\right\}\equiv
\frac{\delta A}{\delta q_i}\frac{\delta B}{\delta \Pi^i}-
\frac{\delta A}{\delta \Pi_i}\frac{\delta B}{\delta q^i},
\label{poissonbracket}
\ee
where $i$ which labels each field is summed;
the dirac matrix is defined by
\be
C_{\iota\kappa}\equiv\left\{\phi_\iota,\phi_\kappa\right\},
\label{diracm}
\ee
and the dirac bracket is defined by
\be
\left\{A,B\right\}\ast\equiv\left\{A,B\right\}-
\left\{A,\phi_\iota\right\}{\rm Inv}\left(C_{\iota\kappa}\right)\left\{\phi_\kappa,B\right\},
\label{diracb}
\ee
where Inv$(C_{\iota\kappa})$ denotes the inverse of $C_{\iota\kappa}$.
For a perfect fluid the constraints are
\be
\phi_1=\Pi^s-\theta \Pi^\sigma,~~~
\phi_2=\Pi^\theta,
\label{pfcons}
\ee
and the dirac matrix \ref{diracm} becomes
\be
C_{\iota\kappa}=-\imath\sigma_{p2}\Pi^\sigma,~~~
{\rm Inv}C_{\iota\kappa}=+\imath\frac{\sigma_{p2}}{\Pi^\sigma},
\label{pfdm}
\ee
where $\sigma_{p2}$ is the pauli matrix
\ber
\sigma_{p1}\equiv
\left(\begin{array}{cc}
    0  & 1  \\
    1  & 0
\end{array}   \right),~~~
\sigma_{p2}\equiv
\left(\begin{array}{cc}
    0  & -\imath  \\
    +\imath & 0
\end{array}   \right),~~~
\sigma_{p3}\equiv
\left(\begin{array}{cc}
    1  &  0  \\
    0  & -1
\end{array}   \right).
\label{paulimatrix}
\ear
The dirac bracket \ref{diracb} for \ref{pfcons} is
\be
\{A,B\}\ast=\{A,B\}+\frac{1}{\Pi^\sigma}\frac{\delta B}{\delta \theta}
\left(\frac{\delta A}{\delta s}-\theta\frac{\delta A}{\delta \sigma}
+\Pi^\sigma\frac{\delta A}{\delta \Pi^\theta}\right)
-A\leftrightarrow B,
\label{pfdb}
\ee
acting on the clebsch momenta and clebsch potentials
\ber
&&\{\Pi^\sigma,\sigma\}\ast=-1,~~~
\{\Pi^\theta,\theta\}\ast=0,~~~
\{\Pi^s,s\}\ast=-1,\n
&&\Pi^\sigma\{\sigma,\theta\}\ast=-\theta,~~~
\Pi^\sigma\{\theta,s\}\ast=-1,~~~
\Pi^\sigma\{\sigma,s\}\ast=0,
\label{pfcommun}
\ear
replacing the dirac brackets by quantum commutators there does not seem to be an easy
representation of the resulting algebra as discussed in \cite{mdr27}.
To quantize one replaces the dirac bracket with a quantum commutator
\be
\left\{A,B\right\}\ast
\rightarrow\frac{1}{\imath\hbar}\left[\hat{A}\hat{B}-\hat{B}\hat{A}\right],
\label{qdb}
\ee
in the present case the second equation of \ref{pfcommun} suggests that $\Pi^\theta=0$,
which is assumed from now on.
There is the problem of how to realize the coordinate commutation relations
in the last three of \ref{pfcommun}
which does not seem possible using the pauli matrices \ref{paulimatrix}
and so this is left for now.
Suppose that $\hat{\Pi}^\sigma$ is replaced by a differential operator
\be
\hat{\Pi}^\sigma=-\imath\hbar\p_x,
\label{diffop}
\ee
possibilities for $x$ include spacetime coordinates $x_a$,
the particle number $n$ or the proper time $\tau$ which restrict $\sigma$
as to a lesser extent does the dirac operator $\gamma^a\p_a$
so the simplest choice is taken $\p_\sigma$.
To proceed it is necessary to have an explicit hamiltonian which does not exist in the
present case as the pressure $\pr$ and the density $\mu$
are not directly expressible in terms of the clebsch potentials.
\section{The explicit linear lagrangian}\label{eospf}
The lagrangian is taken to be linear in $W_a^2$
\be
\Lg=-\frac{1}{2}{\cal F}(n,q)W_a^2-\mho(n,q),
\label{feqs}
\ee
where ${\cal F}(n,q)$ and $\mho(n,q)$ are the first and zeroth order coefficient
functions of the particle number and clebsch potentials respectively.
Metric variation yields
\be
T_{ab}={\cal F}W_aW_b+\Lg g_{ab}={\cal F}h^2V_aV_b+\Lg g_{ab},
\label{fmet}
\ee
requiring this stress to be that of a perfect fluid places the restriction
\be
{\cal F}h^2=nh=\pr+\mu,~~~{\rm or}~~~{\cal F}=\frac{n}{h},
\label{noverh}
\ee
again requiring that the pressure is the lagrangian and using $W^2_a=-h^2$ gives
\be
\Lg=\pr=\frac{1}{2}nh-\mho=\frac{1}{2}\left(\pr+\mu\right)-\mho,
\label{nlg}
\ee
which yields the linear equation of state
\be
\pr=\mu-2\mho,
\label{eosl}
\ee
this is a very restrictive equation of state as it does not have even simple
cases such as $\pr=(\gamma-1)\mu$ as examples.
Variation with respect to the clebsch potentials gives
\be
\dot{n}+n\Theta-\bar{\mho}_{,\sigma}=0,~~~
n\dot{s}+\bar{\mho}_{,\theta}=0,~~~
n\dot{\theta}-\bar{\mho}_{,s}+\theta\bar{\mho}_{,\sigma}=0,
\label{eaemo}
\ee
where
\be
\bar{\mho}_q=\mho_{,q}-\frac{1}{2}h^2F_{,q},
\label{barmho}
\ee
and \ref{noverh} has been used.
Using the equations of motion \ref{eaemo} the bianchi identities \ref{bid} become
\be
T^{ab}_{..;b}=\frac{1}{2}h^2{\cal F}_a
-\left(\theta\bar{\mho}_\theta+\frac{1}{2}h^2{\cal F}_s\right)s_a
-n\theta\left(\frac{1}{n}\bar{\mho}_\theta\right)_a
-\sigma_a\mho_\sigma
-\theta_a\mho_\theta,
\label{lbid}
\ee
for the thermodynamical case \ref{cbemo} and \ref{eaemo} give
zeroth order coefficient function $\mho=nTs$
and then the bianchi identity reduces to ${\cal F}_a=0$ or $n=kh$,
thus what in lagrangian \ref{feqs} looked like an arbitrary function ${\cal F}$ has been
reduced to a constant and this will be further considered in the next section.
Variation with respect to everything else except the particle number $n$
gives the same as the implicit approach of the previous section \S\ref{pf}.
For variation with respect to the particle number $n$ there are three choices:
the {\it first} is to simply do it in which case,
for lagrangians in which $n$ is separated and linear,
the vectors $W,~V$ are forced to be null
and the system is no longer that of a perfect fluid,
the {\it second} is to ignore variation with respect to $n$,
for lagrangians in which $n$ is separated and linear this amounts to assuming a first
law of the form \ref{1law} and one is essentially working in the implicit formalism
of the previous section \S\ref{pf},
the {\it third} is to alter lagrangians with coefficient functions ${\cal F},\mho$
that depend on $n$,
then variation with respect to $n$ can be chosen so that $h^2{\cal F}_{,n}=-2\mho_{,n}$.
Using \ref{pfmom} for $n$,  \ref{shpf} for $h$ and \ref{simplecleb} for $V_a^2$
the lagrangian is
\be
\Lg=\pr=\frac{1}{2}\dot{\sigma}\Pi^\sigma-\mho,
\label{elag}
\ee
using \ref{eaemo} the hamiltonian is
\be
\Hm=\frac{1}{2}\dot{\sigma}\Pi^\sigma+\mho-\frac{\theta}{n}\Pi^\sigma\mho_{,\theta},
\label{eham}
\ee
when $\mho_{,\theta}=0$ the hamiltonian equals the density $\Hm=\mu$.
For this explicit form \ref{eham}
it is possible to use the operator substitution \ref{diffop} so that
\be
\Hm\Psi=-\frac{1}{2}\imath\hbar\dot{\sigma}\Psi_\sigma+\mho\Psi=0,
\label{hschro}
\ee
taking $\dot{\sigma}\Psi_\sigma=\dot{\Psi}$, \ref{hschro} becomes
\be
\imath\hbar\dot{\Psi}=2\mho\Psi,
\label{odepsi}
\ee
integrating
\be
\Psi=A\exp\left(-\frac{2\imath}{\hbar}\int\mho d\tau\right),
\label{wavefn}
\ee
this wavefunction turns out to be too restrictive to be of any use.
\section{Simplest three clebsch potential fluid}\label{tpoq}
The explicit linear thermodynamic lagrangian is
\be
\Lg=-\frac{1}{2}W_a^2-nTs,
\label{tol}
\ee
see the remarks after \ref{lbid}.  Varying with respect to the potentials gives
\be
W^a_a=\bx\sigma+\theta_a s^a+\theta\bx s=0,~
s^a\left(\sigma_a+\theta s_a\right)=0,~
\theta^a\left(\sigma_a+\theta s_a\right)=hT.
\label{teom}
\ee
If these equations are thought of as functions of one variable then the second equation gives
either $s_a=0$ or $W_a=0$ both of which are of no practical use:  therefore two variables are
used specifically to seek plane wave solutions.
A simple choice is
\be
\sigma=A\exp\imath\left(k_0t+k_1x\right),~
\theta=B\exp a\imath\left(k_0t+k_1x\right),~
s=C\exp(1-a)\imath\left(k_0t+k_1x\right),
\label{wave}
\ee
which gives
\be
W_a=\imath\left[k_0,k_1,0,0\right]{\cal K},~~~~~~~~
{\cal K}\equiv\left(\frac{A}{BC}+1-a\right)\theta s,
\label{WK}
\ee
the first and second equations of motion \ref{teom} give
\be
(k_0^2-k_1^2){\cal K}=0,
\label{12eom}
\ee
and the third gives
\be
(k_0^2-k_1^2){\cal K}=\frac{T}{a\theta},
\label{3eom}
\ee
thus the system is decoupled,  in other words as \ref{12eom} forces it to be null
the terms in the first equation of \ref{teom} vanish separately,
and by \ref{3eom} the system is at zero temperature.
Working through in the same manner with the clebsch potentials having both left and right movers,
i.e. like \ref{mode},
the same problems arise and there is no indication that mixed movers
can generate non-zero temperature.
Working through without the restriction that the clebsch potentials are co-directional produces
equations too general to proceed with.

The hamiltonian is
\be
\Hm=-\frac{1}{2}\dot{\sigma}\Pi^\sigma-Ts\Pi^\sigma,
\label{h3}
\ee
in the present case $\dot{\sigma}=h=n=-\Pi^\sigma$ so that
\be
\Hm=\frac{1}{2}\Pi^\sigma-Ts\Pi^\sigma
\label{h3t}
\ee
using the substitution \ref{diffop} with $x=\sigma$ gives
\be
\Hm\Psi=-\frac{\imath\hbar}{2}\Psi_{\sigma\sigma}+Ts\Psi_\sigma=0,
\label{hw3}
\ee
which has solution
\be
\Psi=A\sigma\exp\left(-\frac{2\imath Ts}{\hbar}\right)+B,
\label{psi3}
\ee
which is again too restrictive.
\section{One clebsch potential fluid with two coefficient functions}\label{opof}
The lagrangian is taken to be
\be
\Lg=-\frac{1}{2}{\cal F}(\sigma)\sigma_a^2-\mho(\sigma),
\label{1potlag}
\ee
the metric stress is
\be
T_{ab}={\cal F}\sigma_a\sigma_b+\Lg g_{ab},
\label{1potms}
\ee
$\pr,\mu,n,h$ are recovered in the same way as in the last section.
Varying with respect to $\sigma$
\be
{\cal F}\bx\sigma+{\cal F}^a\sigma_a+\mho_\sigma=0,
\label{1potwe}
\ee
which for simple ${\cal F}$ has simple spherically symmetric solutions;
for wave solutions use zeroth order coefficient function $\mho=m^2\sigma^2/2$ and
\be
\sigma_\pm=A_+\exp(\imath k\cdot x)\pm A_-\exp(-\imath k\cdot x),
\label{mode}
\ee
with $\sigma=\sigma_+$, \ref{1potwe} becomes
\be
\left({\cal F}k_a^2+m^2\right)\sigma+{\cal F}'k_a^2\sigma_-^2=0,
\label{wesol}
\ee
the last term forces either $A_+$ or $A_-$ to vanish.
\section{One fluid described using one vector field}\label{ofove}
In the one fluid one vector approach one hopes to find a lagrangian which recovers as much of
the stress \cite{MTW}eq.22.16d as possible
\ber
T_{ab}&=&(\mu+\pr-\xi\Theta)U_aU_b-2\eta\st{U}{\sigma}_{ab}+2q_{(a}U_{b)}+(\pr-\xi\Theta)g_{ab},\n
S_a&=&nsU_a+\frac{q_a}{T},
\label{mtwfluidstress}
\ear
where $\mu$ is the density,  $\pr$  is the pressure,  $\xi\ge0$ is the coefficient of bulk viscosity,
$\eta\ge0$ is the coefficient of dynamic viscosity,  $\st{U}{\sigma}$ is the shear,  $q$ is the heat flux,
$S_a$ and $s$ are the entropy vector and scalar and $T$ is the temperature.
Consider a fluid lagrangian dependent on one velocity which can be expanded
\be
\Lg=\Lg(V)=\Lg(V^{0+},V^1,V^{1+},V^2,\ldots)
\label{lagexpan}
\ee
where
\ber
\Lg(V^{0+})&=&k^{0+}\pr,\\
\Lg(V^1)&=&k^1\Theta=k^1V^a_{.;a},\n
\Lg(V^{1+})&=&k^{1+}\pr\Theta,\n
\Lg(V^2)&=&k^2_1\dot{\Theta}+k^2_2R_{ab}V^aV^b+k^2_3\omega^2+k^2_4\sigma^2+k^2_5\Theta^2+k^2_6\dot{V}^a_{.;a}\n
&\ddots&\nonumber
\label{lagterms}
\ear
so that the integer superscript on the velocity $V$ indicates the power in which it occurs in the lagrangian,
the meaning of the superscript $+$ will become apparent later.
Here just the first three terms are considered,  terms of second or Raychadhuri order and higher are ignored,
as are any auxiliary,  image,  entropy or electromagnetic fields.
From a technical point of view just the first term in this expansion has been considered in
the previous section \S\ref{pf} and in this section the next two terms are considered.
Metrical variation gives the stress
\be
T_{ab}=(\mu+\pr)\left(k^{0+}+k^{1+}\Theta\right)V_aV_b-2k^1V_{(a;b)}+\Lg g_{ab},
\label{metstress3}
\ee
\ref{metstress3} does not bare much resemblance to \ref{mtwfluidstress} except that there is
a term similar to bulk viscosity appearing.   Variation with respect to the clebsch potentials,
particularly of $\Lg(V^1)$,  gives long expressions which are not helpful.
Variation with respect to the clebsch velocities gives the momenta
\be
\Pi^\phi=-n\left(k^{0+}+k^{1+}\Theta\right),~~~
\Pi^s=\theta\Pi^\phi,~~~
\Pi^\theta=0,
\label{oomom}
\ee
and these give two constraints
\be
\phi_1=\Pi^s-\theta\Pi^\phi,~~~
\phi_2=\Pi^\theta.
\label{ofovc}
\ee
In the present case, as the constraints \ref{ofovc} are the same as for the perfect fluid,
the dirac bracket between the coordinates and momenta and thus the quantum relations
are the same as for the perfect fluid as given above and in \cite{mdr27}.
\section{One fluid described using two vector fields}\label{oftve}
If one tries to incorporate entropy by choosing it to move in the same direction as fluid flow then
the entropy is proportional to the enthalpy and $s=kh$ and the first law becomes
\ber
\delta p&=&n\delta h-nT\delta s
=-nV_a\delta W^a-nTk\delta h\n
&=&-nV_a\delta W^a+nTkV_a\delta W^a
=-n^*V_a\delta W^a,\n
&&{\rm where~~~}n^*\equiv(1-kT)n,
\label{coment}
\ear
so the problem becomes the same as before with the particle number $n$ replaced by $n^*$.
In the one fluid two vectors approach one proceeds as before
except the clebcsh decomposition \ref{simplecleb} is replaced by
\ber
&hV_a=W_a=\sigma_a+\eta\chi_a,&V_aV^a=-1,\n
&\ell U_a=X_a=\alpha_a+\beta\gamma_a,&U_aU^a=-1,
\label{oftvc}
\ear
and the first law \ref{1law} is replaced by
\be
\delta p=n\delta h-nT\delta S=-nV^a\delta W_a+nTU^a\delta X_a.
\label{1law2}
\ee
Metrical variation gives
\be
T_{ab}=nhV_aV_b-nT\ell U_aU_b+\pr g_{ab},
\label{oftvmv}
\ee
when either $T ~ {\rm or} ~ \ell \rightarrow 0$ the second term vanishes.
\ref{oftvmv} can be re-written as
\be
T_{ab}=(\pr+\mu)\left(V_aV_b-\frac{T}{h \ell}X_aX_b\right)+\pr g_{ab},
\label{oftvstress}
\ee
Two choices {\it firstly} the free case where the clebsch potential are all independent,
{\it secondly} the thermodynamic case where the clebsh potential are chosen so that
the thermasy and entropy change are non-vanishing.
The free case essentially duplicates the one vector case of the perfect fluid \S\ref{pf}.
For the thermodynamic case choose
\be
\theta=\eta=\beta,~~~~~~~
s=\chi=\gamma,
\label{thermopot}
\ee
in \ref{oftvc}.
Defining two absolute derivatives
\be
\dot{T}_{abc\dots}=V^eT_{abc\dots;e},~~~
T'_{abc\dots}=U^eT_{abc\dots;e},
\label{defabsol}
\ee
variation with respect to $\sigma,\alpha,\theta{\rm~and~}s$ gives
\be
\dot{n}+n\st{V}{\Theta}=0,~~~~~
(nT)'+nT\st{U}{\Theta}=0,~~~~~~~
\dot{s}=Ts',~~~~~
\dot{\theta}=T\theta',
\label{eqmopot}
\ee
respectively.
The relationship between dot and dash derivatives is needed,  note
\be
X_a=(\alpha-\sigma)_a+W_a,
\label{relXW}
\ee
thus for an arbitrary tensor $A_{abc\dots}$ there is the relationship between the derivatives
\be
\ell A'_{abc\dots}=(\alpha-\sigma)^eA_{abc\dots;e}+h\dot{A}_{abc\dots},
\label{reldd}
\ee
using \ref{reldd}, \ref{eqmopot} becomes
\ber
&&h\st{V}{\Theta}-\ell\st{U}{\Theta}=\left(\ln(nT)\right)^a(\alpha-\sigma)_a+h\left(\ln(T)\right)^\circ,\n
&&\dot{s}=\frac{\ell(\alpha-\sigma)^as_a}{\ell-Th},~~~~~~~
\dot{\theta}=\frac{\ell(\alpha-\sigma)^a\theta_a}{\ell-Th}.
\label{sthetadot}
\ear
In the limit that the two vectors coincide $\alpha\rightarrow\sigma$ and \ref{relXW} and \ref{sthetadot} give
$\dot{T}=\dot{s}=\dot{\theta}=0$ so that the thermasy relation is not recovered unless $T=0$.
The dot momenta are the same as that given by \ref{pfmom},  the dashed momenta are
\be
\Pi^\alpha({\rm dashed})=nT,~~~
\Pi^s({\rm dashed})=n\theta T,
\label{dmom}
\ee
it is necessary to convert these to dot momenta using
\be
\Pi^i\equiv\frac{\delta I}{\delta \dot{q}^i}
=\frac{\delta q'^i}{\delta \dot{q}^i}\frac{\delta I}{\delta q'^i}
=\frac{\delta q'^i}{\delta \dot{q}^i}\Pi^i({\rm dashed})
\equiv f(q)\Pi^i({\rm dashed}),
\label{ddashed}
\ee
where $i$ is not summed and $f$ is an undetermined function of the clebsch potentials;
collecting together \ref{pfmom},\ref{dmom},\ref{ddashed} gives the total momenta
\be
\Pi^\sigma=-n,~~~
\Pi^\alpha=nTf,~~~
\Pi^\theta=0,~~~
\Pi^s=-n\theta+n\theta Tf,
\label{totmom}
\ee
and the three constraints
\be
\phi_1=\Pi^s-\theta\Pi^\sigma-\theta\Pi^\alpha,~~~
\phi_2=\Pi^\theta,~~~
\phi_3=\Pi^\alpha+Tf\Pi^\sigma,
\label{totcons}
\ee
the dirac matrix is
\be
C_{12}=-\Pi^\sigma-\Pi^\alpha,
C_{13}=\left(-\frac{\delta f}{\delta s}
+\theta\frac{\delta f}{\delta \sigma}
+\theta\frac{\delta f}{\delta \alpha}\right)T\Pi^\sigma,
C_{23}=-\frac{\delta f}{\delta \theta}T\Pi^\sigma,
\label{totdm}
\ee
the inverse Inv$C$ can be found,  but there is no simplification in further expressions.
\section{Conclusion}\label{conc}
In \S\ref{pf} the lagrangian approach to perfect fluids was presented,
the clebsch potentials contain more information than is needed to describe the system
so that it is constrained,  dirac constraint analysis removes the superfluous degrees of
freedom resulting in the algebra \ref{pfcommun},  see also \cite{mdr27},  which when quantized
via \ref{qdb} does not seem to lead to a recognizable algebra;  also as the pressure and density
are implicit rather than explicit functions of the clebsch potentials it is not clear what
quantum operators should represent them.
In \S\ref{eospf} to overcome lack of explicit expressions for the pressure and density an explicit
lagrangian \ref{feqs} was studied,  this has a very restrictive equation of state \ref{eosl},
and for thermodynamic choices of the functions reduces to the example of the next section.
In \S\ref{tpoq} the simplest three potential lagrangian \ref{tol} was studied,
it seems to lead to an unrealistic quantum theory \ref{hw3}.
In \S\ref{opof} a one potential lagrangian \ref{hw3} was studied,
it is just a simple generalization of the klein gordon lagrangian,
but the generalization is obstructive enough to prevent successful
investigation of it by the fourier oscillator method.
In \S\ref{ofove} the perfect fluid is generalized so that
it depends on higher powers of the comoving fluid,
a term similar to the bulk viscosity appears
and the momentum constraints are the same as those for a perfect fluid.
In \S\ref{oftve} the perfect fluid is generalized to include two vector fields in the hope that
these can represent both density and entropy flow,
equating some of the clebsch potentials in each vector leads
to plausible thermodynamic equations \ref{sthetadot}.
\section{Acknowledgements}\label{acknowledgements}
I would like to thank Tom Bridges, Alex Craik and Mattias Marklund for their interest
in parts of this work.

\end{document}